\documentstyle[aps,prb,amssymb,float,graphicx,twocolumn]{revtex}
\begin{document}
\draft

\wideabs{
\title{Coupling of the lattice and superlattice deformations and
hysteresis in thermal expansion for the quasi one-dimensional
conductor TaS$_3$
}
\author{A.V.~Golovnya, V.~Ya.~Pokrovskii, and P.M.~Shadrin}
\address{Institute of Radioengineering and Electronics,
Russian Academy of
Sciences, 103907 Moscow, Russia}
\date{\today}
\maketitle
\begin{abstract}
An original interferometer-based setup for measurements of length
of needle-like samples is developed, and thermal expansion of
o-TaS$_3$ crystals is studied. Below the Peierls transition the
temperature hysteresis of length $L$ is observed, the width of the
hysteresis loop $\delta L/L$ being up to  $5 \cdot 10^{-5}$. The
behavior of the loop is anomalous: the length changes so that it is
in front of its equilibrium value. The hysteresis loop couples
with that of conductivity. The sign and the value of the length
hysteresis are consistent with the strain dependence of the
charge-density waves (CDW) wave vector.  With lowering temperature
down to 100~K the CDW elastic modulus grows achieving a value
comparable with the lattice Young modulus. 
Our results
could be helpful in consideration
of different systems with intrinsic superstructures.
\end{abstract}

\pacs{PACS Numbers: 71.45.Lr, 65.40.De}}

Internal degrees of freedom is a feature of a random system;
in principle, they can give rise to metastable size states resulting, say, to 
hysteresis in thermal expansion.\cite{Nagel}
A special class form the compounds with intrinsic 
superstructures. 
Comprising two periodicities, generally incommensurate, 
the compounds occupy intermediate place between genuine 
aperiodic and truly periodic systems.\cite{Voit} In these systems, 
such as charge- and spin-density waves (CDW and SDW),
\cite{Bribzor} Wigner cristals, 
superconductors in magnetic 
fields,\cite{LTobzor} structurally incommensurate crystal phases,
\cite{INCOMobzor}
the superstructure periodicities could be varied by external fields
or temperature changes.
The resulting metastable configurations  
can be reflected back onto the elastic properties and size 
of the underlying lattice,\cite{Bribzor,LTobzor,INCOMobzor}
though this question is still poorly understood.

Quasi 1-dimensional conductors with CDW belong to a widely studied
class of materials,
in which intrinsic superstructure develops
through the Peierls transition \cite{Grun}.
When electrons condense into CDW they form a deformable medium, --
an electronic crystal.
Deformation of the CDW affects their main static and dynamic
properties and gives rise to 
metastability and hysteresis. 

The straightforward treatment of the CDW as a spring, whose strain is just
applied to the crystal at the ends or via the impurities is not valid.
Moreover, in the simple one-dimensional model the strains of the CDW and the
crystal do not couple at all: 
if initially the CDW are relaxed, any change
of the crystal length would not draw the CDW away from the
equilibrium, {\it i.e.} give rise to a CDW deformation,\cite{1dmodel} 
as it was noticed in Refs.~ \onlinecite{Mozurkth,Brill1,Mozurkexp,Brill2}.
Similarly, once the CDW is deformed, any change of the lattice
constant, $c$, would neither decrease nor increase the deviation of
the CDW wavelength $\lambda$ from the equilibrium value, $\lambda_{eq}$. 
So, within this model a CDW
deformation would not give rise to a length change.

At the same time, the interaction of the CDW and the lattice is
clearly seen from the elastic anomalies, including a drop of the
Young modulus of the lattice,
\cite{Mozurkth,Brill1,Mozurkexp,Brill2} $Y_{l}$, up to
4\%,\cite{Mozurkexp} when the CDW become depinned. 
Mozurkewich \cite{Mozurkth} has concluded that the lattice deformation
{\it does} give rise to a deformation of the CDW: when the CDW are
at rest, they cannot relax, and so they contribute to the
total elastic energy; the sliding CDW relax rapidly and
their contribution drops out.\cite{Mozurkth} For his model
Mozurkewich \cite{Mozurkth} introduced an empirical parameter $g$
reflecting the deviation of $\lambda _{eq}$ from the simple
one-dimensional model:

\begin{equation}
\delta \lambda _{eq}(c)/\lambda _{eq}(c_{eq})=
(g+1)\delta c/c_{eq}, 
\label{lambdaeq}
\end{equation}
$g\neq 0$. However, no way was proposed to estimate $g$.

Another remarkable observation was reported in
Ref.~\onlinecite{HBZI}: the length $L$ of TaS$_3$ samples as a
function of electric field demonstrated hysteresis partly scaling
with that of resistance. The values of $L$ obtained at different
directions of the voltage sweep differed by $\sim 10^{-6}$. This
result was also treated in terms of coupling of the CDW strains
with the deformation of the pristine lattice. 
Note that the
field-induced length change (as well as the change of resistance)
results only from an inhomogeneity of the sample properties:
obviously, the length could depend on the voltage polarity only if
the inversion symmetry of the sample is broken \cite{HBZI}.
Thus, study of the {\it electric-field} induced hysteresis of length
can not provide complete understanding of the CDW-lattice
coupling.

To study the effect of the CDW deformation on the crystal length
it could be more fruitful to observe {\it thermal} hysteresis of
$L$. With changing temperature the $q$-vector ($q
\equiv 2\pi/\lambda$) falls {\it behind} its equilibrium value, which is
temperature dependent;\cite{q(T)} the deviation of $q$ from the
equilibrium is limited by the critical deformation at which the 
phase slippage (PS) begins.\cite{critstat} So, thermal
cycling creates a CDW deformation, which is relatively uniform
along the sample and is of the maximum possible value.

For our study we chose o-TaS$_3$ as a representative CDW
conductor. Among the quasi one-dimensional conductors TaS$_3$ is
one of the most widely studied, including the elastic properties.
TaS$_3$ demonstrates the Peierls transition at $T_P=220$~K, below
which the resistance follows an activation law revealing the
half-gap $\Delta=700$~K. The dependence $R(T)$ demonstrates a
pronounced hysteresis, the width of the loop $\delta R/R$ being up
to 50~\% at $T$ around 100~K.

In the present Letter we report temperature hysteresis of length
for o-TaS$_3$; the hysteresis couples with that of resistance 
having an anomalous sign. 
A quantitative treatment of the
effect is proposed.

\begin{figure}[h]
\hskip -.635cm
\includegraphics[scale=.5]{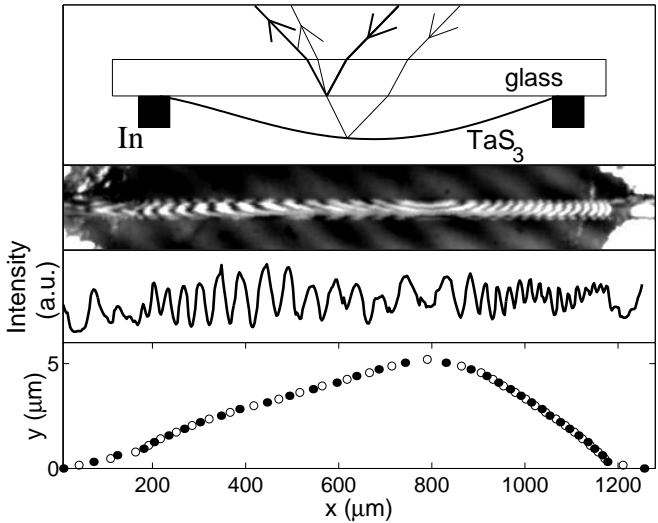}
\\
\caption{a) Arrangement of a TaS$_3$ sample on the substrate and
the scheme of the reflections and interference of the laser beams
(the incidence is close to normal). b) An example of
an image obtained, c) -- the section of the image,
d) -- the resulting profile of the sample. The open circles
correspond to the minima in the section (Fig.1c), and the closed
circles --- to the maxima.} 
\end{figure}

We studied samples of TaS$_3$ with typical length 1~mm,  width
15--35~$ \mu$m, and thickness 5-10 times less. 
The samples were arranged on a transparent
glass substrate  (Fig.~1a). The contacts
were fixed on the substrate with indium, while the central part
of the sample formed an arc. 
The inner surface of the substrate played the role
of a semi-transparent mirror. The laser beam with wavelength
$\lambda_l=635$~nm fell down through the substrate, and 
partially reflected from the inner surface of the substrate and
from the sample surface forming an interference pattern (similar
with the wedge interference). The pattern was fixed with a video
camera combined with a microscope. An example of such a pattern is
shown in Fig.~1b. The neighboring dark or bright fringes
correspond with change of the sample distance $y$ from the
substrate surface by $\pm \lambda_l/2$. Making sections of such
images (an example is shown in Fig.~1c) we obtained the profiles
$y(x)$ of the samples, like the curve presented in Fig.~1d.
Finally, the length of such curves was calculated. We 
detect a length change about 5~\AA~ which corresponds to the
relative length change $\delta L/L=5\cdot10^{-7}$, though in
principle the sensitivity could be improved.\cite{vasuki} 
Our estimates have shown that the
contributions to $\delta L$ due to the changes of stress and $Y_l$
are negligible.\cite {elsewhere} Thus, the observed changes of $L$
with temperature 
were associated with the change
of the equilibrium sample length with respect to the substrate. 
To have absolute results we have measured 
thermal expansion of the glass substrate in addition.

\begin{figure}[h]
\hskip -.1cm
\includegraphics[scale=.47]{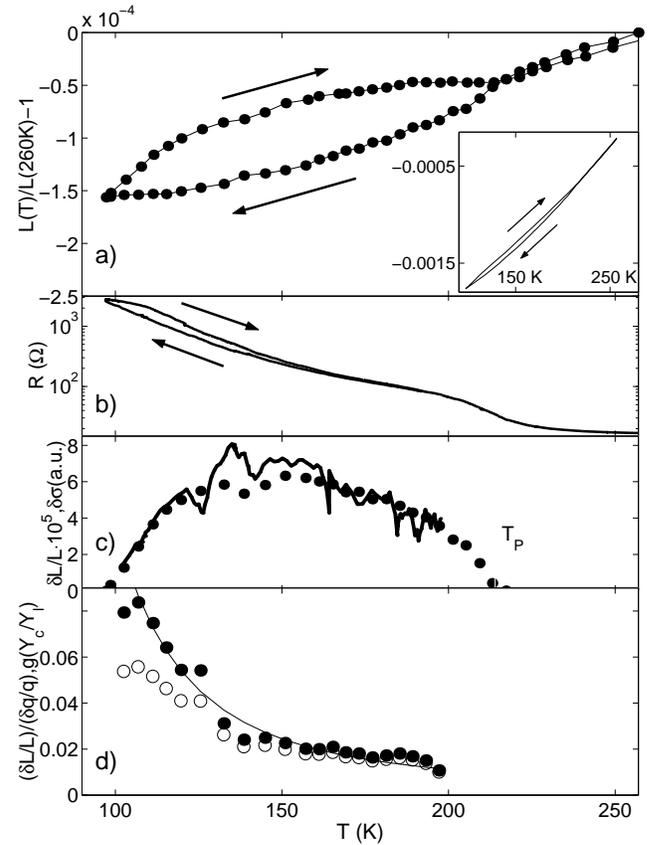}
\\
\caption{a) Temperature dependence of $\delta L/\L$ with
respect to glass. The absolute thermal expansion is shown in the
{\it inset}. b) $R(T)$ measured simultaneously. c)
The width of the length (the circles) and conductivity (the solid
line) hysteresis loops as a function of $T$. d) The ratio
$(\protect\delta L/L)/(\protect\delta \protect\lambda /
\protect\lambda)$ (open circles) and $gY_c/Y_l$ (black circles)
resulting from Fig.~2c. We use Eq.~\protect\ref{result} and the
$\protect\mu(T)$ dependence.\protect\cite{mulat}
The solid line indicates  $Y_c \propto \exp(470/T)$.} \label{Fig3}
\end{figure}

Figs.~2a) and b) show the temperature dependences of length (with
respect to glass) and of the resistance for the representative
sample of TaS$_3$. In the inset to Fig.~2a) the absolute
dependence of length is shown.\cite{Canada}
The dependences of length and resistance both
clearly show hysteresis below the Peierls transition temperature
$T_P = 214$ ~K. The length hysteresis loop opens immediately below
$T_P$. The development of hysteresis is the dominating effect near
$T_P$, on whose background it is difficult to distinguish a feature
coupled to the transition in itself.\cite{CanadaTp}
The length
hysteresis grows below $T_P$ achieving $\delta L/L = 5 \cdot
10^{-5}$ (Fig.~2a), which by 1.5 orders of magnitude exceeds the
maximum value reported in Ref.~\onlinecite{HBZI}. At the same
temperature the states approached from lower temperature have
higher length than those approached from above. Correspondingly,
application of electric field exceeding the threshold reduces the
length after heating the sample, and increases after cooling.
Thus, the main effect of electric field is the relaxation of
the thermally induced states. The
length change induced by the field in itself was  $\sim 10^{-6}$, 
like in Ref.~\onlinecite{HBZI}. 
Note, that the sign of the hysteresis is non
typical: the length is {\it in front} of its equilibrium value (inset to
Fig. 2a), but not behind.

The length hysteresis apparently couples with the resistance hysteresis,
higher length corresponding to higher resistance at a given 
temperature.\cite{antiHBZI} The
coupling between $R$ and $L$ is well seen from Fig.~2c, where the
temperarure dependence of $\delta L/L$ is presented together with the loop
of conductivity, $\delta \sigma \equiv 1/R_{cool}-1/R_{heat}$. Note that in
the unipolar approximation 
$\delta \sigma \approx \sigma_{300}(\delta \lambda/\lambda)(\mu/\mu_{300})$,
\cite{bzn} where the index
``300'' marks the room--temperature values. So, the $\sigma$ hysteresis
reflects that of the CDW wavelength to the accuracy of the temperature
dependence of the mobility $\mu$, which in turn can be found from 
Ref.~\onlinecite{mulat}. 
Fig.~2d shows the ratio $(\delta L/L)/(\delta \lambda
/\lambda)$ {\it vs.} $T$. 
At the low temperatures 
$\delta L/L$ is $\sim 6$~\% of the CDW deformation.

As we have noticed in the beginning, the interactions of
the CDW and the sample cannot be presented as that of two springs
connected in parallel (the lower inset to Fig.~3). Though we do
not see a way to obtain quantitatively the strain dependence of
the $q$-vector for TaS$_3$, one can recall experimental
results, from which it is possible to estimate the value of $g$
(Eq.~\ref{lambdaeq}).
It has been found \cite{PT} that at a certain value
of uniaxial strain $S^*$ the properties of the CDW 
change abruptly: {\it e.g.}, the non-linear conduction nearly disappears.
This has been attributed to the transition of the CDW to the
4-fold commensurability.
Comparing the normalized dependences of the $q$-vector \cite{q(T)}
and of the strain $S^*$ \cite{PT}
on temperature (Fig.~3) one can see that they become similar
\cite{PT} if one multiplies the value of $S^*$ by 6. Thus, one can
assume that a deformation
of the lattice induces the change of the equilibrium CDW wavelength 
$\lambda_{eq}(c)$ in accordance with Eq.~(\ref{lambdaeq})
with $g
\approx 6$. Such a large value of $g$ indicates that the simple
1-dimensional model \cite{1dmodel} cannot even roughly describe the
strain-induced change of $\lambda_{eq}$.
Evidently, the change of $\lambda_{eq}$ is dominated by
transverse effects: the longitudinal strain decreases
the thickness of the sample resulting in an
increase of the intrachain coupling. This modifies the form
of the Fermi surfaces, and consequently, the $q$ -vector.\cite{PT}
To comprehend the sign and the value of the $q$ change one should
study the evolution of the Fermi surface in detail. In principle,
we cannot exclude that $g$ is temperature dependent. However,
as we shall see below, the assumption that $g=$~const, namely $g \approx 6$,
is consistent with our experiment.

For the next step,
we present the elastic energy density $W$ as a sum of the lattice and the
CDW energies:
\begin{figure}[h]
\hskip .4cm
\includegraphics[scale=.4]{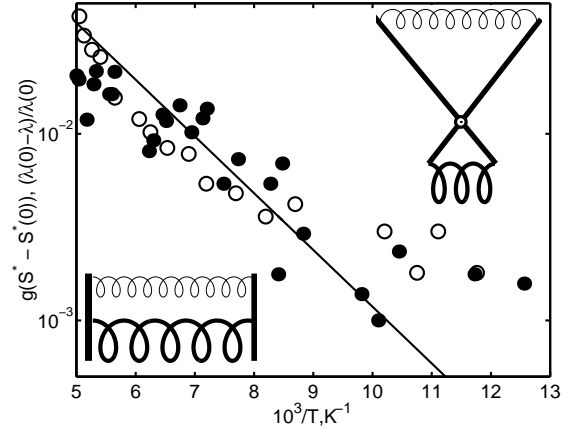}
\\
\caption{The temperature dependences of 
$S^*-S(0)$ (open circles)\protect\cite{PT}
and of the wave vector
\protect\cite{q(T)}  (q-q(0))/q(0))
(closed circles).\protect\cite{S(0)q(0)} $S^*$ is
multiplied by $g=6$. {\it Insets}: the equivalent schemes
illustrating the interaction of the CDW (the thin-line spring) and
the pristine lattice (the thick-line spring). The lower-left
sketch is the naive scheme, which is invalid. The upper-right
sketch illustrates Eq.~\ref{result} ($g > 1$).} \label{Fig1}
\end{figure}

\begin{equation}  \label{gukilton}
 W = \frac{1}{2}(Y_l (\frac{c-c_{eq}}{c_{eq}})^2 + 
Y_c(\frac{\lambda-\lambda_{eq}(c)}
{\lambda_{eq}})^2),
\end{equation}
where $Y_l$ and $Y_c$ are the elastic moduli of the lattice and
the CDW respectively.
Taking into account the condition
(\ref{lambdaeq})\cite{perp} and
assuming that no PS occurs 
($\lambda/c={\rm const}$),
we can minimize $W$ and 
obtain the resulting length
change:

\begin{equation}  \label{result}
\frac{\delta L}{L}=\frac{\delta c}{c_{eq}} = g\frac{Y_c}{Y_l +g^2Y_c }
\frac{\delta \lambda}{\lambda} \approx g\frac{Y_c}{Y_l}\frac{\delta \lambda}
{\lambda},
\end{equation}
where $\delta \lambda$ is the initial CDW deformation (at fixed
$c$). The approximation implies that $g^2Y_c \ll Y_l$.
Eq.~\ref{result} is quite transparent: the crystal deforms as a
spring connected in parallel to the CDW, but with a factor $-g$.
$g>0$ means that, say,  compressed CDW ($\lambda <
\lambda_{eq}$) would result in a decrease of the sample length,
which agrees with our observation. E.g., cooling corresponds to
growth of $\lambda_{eq}$ \cite{q(T)}, so the CDW is in compressed
state ($\lambda-\lambda_{eq} <0$), which results in decreased $L$ (Fig~2a).
The CDW -- crystal interaction could be
illustrated with a scheme consisting of two springs
connected via a lever (Fig.~3, upper inset).

From Eq.~\ref{result} it follows that the ratio $(\delta
L/L)/(\delta \lambda /\lambda)$ (Fig.~2d) equals $gY_c/Y_l$. Note,
however, that with $g=6$, $g^2Y_c$ at low temperatures is
comparable with $Y_l$, and one should use the exact version of
Eq.~\ref{result}. The resulting value of $gY_c/Y_l$ is also shown
in Fig.~2d. One can see the growth of the CDW elastic modulus with
lowering temperature, 
though slower
than $Y_c \propto \exp(\Delta/T)$ \cite{critstat,AV} (see the
solid line in Fig.~2d).\cite{nonun} Note that the curve $(\delta
L/L)/(\delta \lambda /\lambda)$ {\it vs.} $T$ resembles the
temperature dependences of the lattice softening in electric field
\cite{Mozurkexp}. 
This could be expected, since from Eqs.~\ref{lambdaeq}
and \ref{gukilton} it follows that the value $g^2Y_c$ is added to
$Y_l$ when the CDW are at rest, while in the sliding state the CDW
contribution drops out \cite{Mozurkth}. 

From the condition of neutrality \cite{AV} accompanying the CDW
deformation we can obtain for the CDW Young modulus:
$Y_c=\frac{d\zeta}{dq}\frac{q^2}{\pi s}$ \cite{critstat}, where
$\frac{d\zeta}{dq}$ is the derivative of the chemical potential
level by the CDW $q$-vector, and $s$ is the  area per chain.
Substituting this in Eq.~\ref{result}, we obtain:

\begin{equation}  \label{dlzeta}
\frac{\delta L}{L} \approx g\frac{2 \delta \zeta}{\lambda s Y_l}.
\end{equation}
For the unipolar (p-type for TaS$_3$) conduction
\cite{critstat,mulat} 
$\delta \zeta= T\ln(R_{heat}/R_{cool})$.
Substituting into Eq.~\ref{dlzeta} $g=6$, 
$\delta \zeta = 50$~K (for $T=120$~K, Fig.~2b), 
$\lambda=10$~\AA, $s=80$~\AA$^2$ (Ref.~
\onlinecite{mobzor}), $Y_l=380$~GPa \cite{Brill3}, we obtain
$\delta L/L = 3 \cdot 10^{-5}$, in agreement with the experiment
(Fig.~2b). This agreement supports the approach we used to estimate $g$,
and proves that under a certain uniaxial strain ($S^*$) the CDW
in fact transit to the 4-fold commensurability.\cite{PT}

In conclusion, we have observed thermal hysteresis of length for
the quasi one-dimensional conductor TaS$_3$. In the framework of
the model proposed
the observed anomalous sign and value of the hysteresis loop 
are consistent with the 
dependence of the $q$-vector on the lattice strain. The results
could be helpful in understanding the behavior of other systems
with intrinsic superstructures, whose parameters could be rearranged due to the
temperature effects or external fields.

We are thankful to G.P.~Vorob'ev, S.G.~Zybtsev, and S.V.~Zaitsev-Zotov
for help in the
experiment and discussions, to S.N.~Artemenko, A.A.~Sinchenko,
V.B.~Preobrazhenskii, G.V.~Stepanov, J.W. Brill, and M.E.~Itkis
for helpful discussions. We acknowledge
RFBR (02-02-17301,
00-02-22000 CNRS), 
the State programs "Physics of Solid-State nanostructures" and
"Low-Dimensional quantum structures", NWO and INTAS-01-0474.


\begin{references}
\bibitem{CanadaTp}If one takes $L(T) \equiv (L_{cool}(T)+L_{heat}(T))/2$,
he will obtain a decrease of length $\sim 10^{-5}$ associated with
the Peierls transition, in qualitative agreement with 
Ref.~\protect\onlinecite{Canada}.
\bibitem{antiHBZI}  In Ref.~\onlinecite{HBZI} the state with smaller $L$
corresponded to higher $R$. Evidently, both the $R$ and $L$
changes reported \cite{HBZI} result from non-complete compensation
of the contact CDW deformations of opposite signs. In this case
one can obtain an arbitrary relation between the $R$ and $L$
hysteresis dominated by the non-linear effects.
Concerning our experiment, the small difference in the forms of
the loops of $L$ and $R$ (Fig.~2c), as well as the small loop of
$L$ above $T_P$ (Fig.~2a), is evidently within our experimental
error.
\bibitem{S(0)q(0)}We took $S(0)=3.9 \cdot 10^{-3}$, and  $q(0)=0.2503c^*$ 
suggesting that both dependences follow the Arrhenius law (the 
slope of the line corresponds to 700~K, which is close to $\Delta$).
\bibitem{perp}  Eq.~\ref{gukilton} implies that the principal part of the
CDW elastic energy is associated with their longitudinal deformations, while
Eq.~(\protect\ref{lambdaeq})
takes into account the dependence of $\lambda_{eq}$ on all components of the
crystal deformation.
\end{references}
\end{document}